\documentstyle[preprint,aps]{revtex}

\begin{document}

\draft

\title{X-Ray Diffuse Scattering Study on Ionic-Pair Displacement Correlations in Relaxor Lead Magnesium Niobate}

\author{Naohisa Takesue and Yasuhiko Fujii}

\address{Neutron Scattering Laboratory, Institute for Solid State Physics, The University of Tokyo, 106-1 Shirakata, Tokai, Ibaraki 319-1106, Japan.}
\author{Hoydoo You}

\address{Argonne National Laboratory, Materials Science Division, 9700 South Cass Avenue, Argonne, Illinois 60439, USA.}

\maketitle

\begin{abstract}
Ionic-pair equal-time displacement correlations in relaxor lead magnesium niobate, $Pb(Mg_{1/3}Nb_{2/3})O_{3}$, have been investigated at 300, 270, and 240K in terms of an x-ray diffuse scattering technique. Functions of the distinct correlations have been determined quantitatively. The results show the significantly strong rhombohedral-polar correlations regarding Pb-O, Mg/Nb-O, and O-O' pairs. Their spatial distribution at 300K forms an ellipse or a sphere with the diameters of 30-80$\AA$, and the sizes are reduced to 30-40$\AA$ on cooling through local condensation of thermal lattice fluctuations. This observation of local structure in the system proves presence of the polar microregions in the paraelectric state which leads to the dielectric dispersion.
\end{abstract}

\paragraph*{}
$Pb(Mg_{1/3}Nb_{2/3})O_{3}$ (called PMN) is a representative system of relaxors, of a special kind of ferroelectrics exhibiting a diffuse cubic-to-rhombohedral phase transition and dielectric dispersion [1,2] near 270K. The crystal structure, based upon a simple cubic perovskite often represented by $ABO_{3}$, includes two different ionic species at site B, $Mg^{2+}$ and $Nb^{5+}$, which are known to make 1:1 chemical short-range order (SRO) along the $<$111$>$ [1,2]. This local atomic nature has been suggested to cause the unusual phenomena, but a microscopic origin of the mechanism is yet to be fully understood.

The ordering nature of PMN is expected to cause two kinds of lattice distortion. The one kind is of an electrostatic origin. B-site occupation of the two ions gives a bulk-averaged valence of +4, and this value electrically neutralizes the whole system with the other species, $Pb^{2+}$, $O_{1}^{2-}$, $O_{2}^{2-}$, and $O_{3}^{2-}$ (Subscripts of three oxygens mean the cell faces where each species is centered. The faces are perpendicular to x,y, and z,respectively). However, because of the SRO, the valence averaged in the ranges becomes +3.5 and hence $\sim$5+ in the Nb-rich surroundings. This charge misfit is suggested to produce lattice distortion electrostatically [3]. The other kind lies in a mechanical origin, a difference in ionic sizes between the B-site ions [4,5], called size-effect, which also leads to the same consequence as that of the former kind. Both effects can cause local destruction of the cubic symmetry giving the permanent dipoles whose presence enables us to evaluate models previously proposed to explain a mechanism of the relaxor behavior [6,7]. Therefore, it is very important to directly observe local ionic displacements in PMN to obtain experimental evidence of existence of the polar correlations, which motivated us to carry  out a measurement of x-ray diffuse scattering. There are preceding experiments performed based upon the same purpose by means of NMR [8] to investigate the polar {\it time} correlations, and by a transmission electron microscopy (TEM) [9] and inelastic neutron scattering (on PZNT which is the different system) [10] to observe the {\it spatial} correlation. Our x-ray diffuse scattering work will be compensative for these previous studies, giving the lengths and the precise symmetry.

An x-ray (also neutron and electron) intensity scattered by atoms in a crystal, defined as $I(\vec{q}_{1})$ where $\vec{q}_{1}$ is a scattering vector, consists of the components given by an averaged crystal structure and local atomic arrangements, $I_{Bragg}(\vec{q}_{1})$ and $I_{Diff}(\vec{q}_{1})$, respectively. The former gives a Bragg intensity, and the latter a diffuse scattering intensity which is tremendously weaker than the other component. In the current case, only $I_{Diff}(\vec{q}_{1})$ is dealt with, and is further divided into the contributions from the local displacements (including phonons) and the SRO, $I_{Disp}(\vec{q}_{1})$ and $I_{SRO}(\vec{q}_{1})$. For mathematical expression of the two intensities, the simple way is to adopt sums of plane waves [11] which represent the atomic configurational disturbance. For example, the displacement of the species $\gamma$ at the s th unit cell, $u_{s \gamma}$, is written in the following form:
\begin{eqnarray}
u_{s \gamma } & = & \sum_{\vec{k}} \sum_{i} \, u_{\gamma \vec{k} i} \hat{e}_{i} \, exp[{\it i}\vec{k} \cdot (\vec{R}_{s} \, + \, \vec{r}_{\gamma})]\,\,,
\end{eqnarray}
where i is an index of real-space Cartesian coordinates, $u_{\gamma \vec{k}i}$ is the displacement-wave amplitude of wavevector $\vec{k}$ allowed in the first Brillouin zone, $\hat{e}_{i}$ is a unit vector, and $\vec{R}_{s}$+$\vec{r}_{\gamma}$ is the atomic position. The concentration is also given in the same way, and has the form similar to Equation (1). If these plane-wave sums are used for the expression, the intensity equations obtain the simple forms in electron units (e.u.). We show only $I_{Disp}(\vec{q}_{1})$ as below:
\begin{eqnarray}
I_{Disp}(\vec{q}_{1}) & = & N^{2} \, \sum_{\gamma} \sum_{\gamma '} \sum_{i} \sum_{i '} \, f_{\gamma} f^{*}_{\gamma '} \, exp[{\it i}(\vec{k}+\vec{q}_{1}) \cdot (\vec{r}_{\gamma}-\vec{r}_{\gamma '})]{} \nonumber \\
& & {}(\hat{e}_{i} \cdot \vec{q}_{1}) \, (\hat{e}_{i'} \cdot \vec{q}_{1}) \, < u_{\gamma \vec{k} i} \, u_{\gamma ' \vec{k} i'}>\,\delta(\vec{k}+\vec{q}_{1}-2 \pi H_{hkl})\,\,,
\end{eqnarray}
where N is the number of unit cells considered, $f_{\gamma}$ is the x-ray atomic scattering factor [12] including the temperature factor [13], and 2$\pi$$H_{hkl}$ is a Bragg reciprocal lattice vector (remove $(\hat{e}_{i} \cdot \vec{q}_{1})(\hat{e}_{i'} \cdot \vec{q}_{1})$ term to gain the form of $I_{SRO}(\vec{q}_{1})$). A bracket of amplitude product $< u_{\gamma \vec{k} i} \, u_{\gamma ' \vec{k} i'}>$ originates from a fact that the correlations which we observe are spatially averaged, $< \cdots >$. The products are only unknowns in Eq.(2) and their values are unique for all vectors $\vec{q}_{1}$ with unique $\vec{k}$. For PMN, there are 17 unknowns including the SRO terms (The number seems small, but is a result from considerable degeneracy among the terms because of space group Pm3m.). Therefore, if at least 17 total scattering intensities observed for universal $\vec{k}$ are obtained, the unknowns can be determined through linear algebra. The computed values for all allowed $\vec{k}$ provide the ionic-pair equal-time displacement correlation functions according to the following definition:
\begin{eqnarray}
<u_{s \gamma i}\,u_{0 \gamma ' i'}> & = & \sum_{\vec{k}} \, < u_{\gamma \vec{k} i} \, u_{\gamma ' \vec{k} i'}> \, exp[{\it i}\vec{k} \cdot ((\vec{R}_{s}+\vec{r}_{\gamma})-(0+\vec{r}_{\gamma '}))]\,\,,
\end{eqnarray}
where $R_{s'}$ is set at the real space origin, s'=0. A plot of Eq.(3) as a function of $R_{s}$ represents strength and distribution of the $\gamma$-$\gamma$' correlation required for this study.

An x-ray diffuse scattering measurement was performed at room temperature (300K) in the air and below 300K with a DISPLEX cryostat on self-flux-grown single crystal PMN whose size is $\sim$6$\times$4$\times$2$mm^{3}$. A surface normal of the largest face is almost parallel to the [001], and with respect to this face, the bisecting-2$\theta$ diffraction was done on a 4-circle diffractometer (H$\ddot{u}$ber) with a scintillation counter. In this way, x-ray absorption by the sample is independent of $\vec{q}_{1}$. The conventional Cuk$\alpha$ rotating-anode generator (Mac Science) was used and operated at 15kW. X-rays were monochromated by a pyrolytic graphite (002). A preliminary measurement of the Bragg intensity distribution showed that the volume-distribution extent is $\sim$$(0.01)^{3}$ in reciprocal lattice units ($r.l.u.^{3}$). Therefore, minimum $\mid\vec{k}\mid$ was determined as 0.02 r.l.u. to avoid the Bragg contamination. The intensities were collected in reciprocal lattice volume with changing $\vec{q}_{1}$ by increment 0.02 r.l.u. (i.e., N=$50^{3}$). The volume contains about 5000 independent $\vec{q}_{1}$s. The points consist of about 160 $\vec{q}_{1}$ sets, and each set includes at least 17 points. The data acquisition time at each point was 5 minutes, and the complete measurement took about 3 weeks. A factor to convert the observed intensities to those in e.u. was determined from the integrated Bragg intensities of an aluminum standard powder sample [14]. Experimental conditions and means for this measurement are the same as those for the diffuse scattering measurement. The contribution of two kinds of unwanted scattering, air scattering and Compton scattering, was subtracted from the observed intensities through the measurement and the calculation [15], respectively.

Figure 1 (a) is a contour map of the observed e.u. intensities at 300K in the 0th (100) reciprocal lattice plane. A portion of the collected data was used for this mapping. Ridges of the intensities are seen around all Brillouin zone centers along the [0$\bar{1}$1]. The tiny ridges along the [011] also appear around some zones, indicating $\vec{q}_{1}$ dependence on the distribution. This feature essentially agrees with that observed in previous x-ray work [3] which suggested the origin from the lattice distortion making competing interactions with the $<$0$\bar{1}$1$>$ transverse optic phonons. The current results show the significant intensities around the zone centers whose three digits of the indices h, k, and l give the following relationship, h+k+l= even integer. The point is that these scattering intensities are nearly isotropic close to the zone centers. Such zones were expected to exhibit a little to no ANISOTROPIC diffuse scattering in the previous work [3], which is consistent with the present results. Emergence of these nearly isotropic intensities suggests that the optic mode which we are dealing with is not purely optic [16]. The impure component may arise from uncollaborative interactions of the charge misfit with the pure mode. The observed data were analyzed based upon Eq.(2), and $< u_{\gamma \vec{k} i} \, u_{\gamma ' \vec{k} i'}>$ terms were determined quantitatively. The intensities were synthesized using the determined parameters. The recalculated map is represented in Fig. 1 (b), and the figure reproduces the observed data very well giving the R factor $\sim$2$\%$.

Results of the parameter values indicate that the three distinct displacement amplitude products, $< u_{Pb \vec{k} x} \, u_{O1 \vec{k} y}>$, $< u_{B \vec{k} x} \, u_{O1 \vec{k} y}>$, and $< u_{O1 \vec{k} y} \, u_{O2 \vec{k} z}>$, are significant only around $\mid\vec{k}\mid$=0 (species B means the B-site ion, $Mg^{2+}$/$Nb^{5+}$.). Their values are largely negative, positive, and positive quite near the zone center, and oscillate almost in phase with each other with increasing $\mid\vec{k}\mid$. Since their absolute values are 20-$10^{5}$ times bigger than all the others, only these three terms are required to describe the local ionic structure. Its qualitative representation is given below.

The negative values of $< u_{Pb \vec{k} x} \, u_{O1 \vec{k} y}>$ means a product of the shifts of Pb and $O_{1}$ along, e.g., -x and +y, respectively. The shifts are illustrated on the left-hand side in Figure 2 (a). Because of the simple-cubic-perovskite symmetry, $< u_{Pb \vec{k} x} \, u_{O1 \vec{k} y}>$ is equivalent to $< u_{Pb \vec{k} x} \, u_{O1 \vec{k} z}>$. Involving $O_{2}$ and $O_{3}$, the further symmetry consideration indicate more equivalencies such as $< u_{Pb \vec{k} x} \, u_{O1 \vec{k} y}>$=$< u_{Pb \vec{k} y} \, u_{O2 \vec{k} x}>$=$< u_{Pb \vec{k} y} \, u_{O2 \vec{k} z}>$=$< u_{Pb \vec{k} z} \, u_{O3 \vec{k} x}>$=$< u_{Pb \vec{k} z} \, u_{O3 \vec{k} y}>$. Therefore, superposition of all the equivalencies gives the resultant shifts along the [$\bar{1}\bar{1}\bar{1}$] for Pb, the [011] for $O_{1}$, the [101] for $O_{2}$, and the [110] for $O_{3}$. This picture is seen in the otther two figures of Fig. 2 (a. Similary, as shown in the right-hand figure of Fig. 2 (b), positive $< u_{B \vec{k} x} \, u_{O1 \vec{k} y}>$ implies a product of the shifts of B and $O_{1}$ along +x and +y, respectively. The equivalencies can be written as $< u_{B \vec{k} x} \, u_{O1 \vec{k} y}>$=$< u_{B \vec{k} x} \, u_{O1 \vec{k} z}>$=$< u_{B \vec{k} y} \, u_{O2 \vec{k} x}>$=$< u_{B \vec{k} y} \, u_{O2 \vec{k} z}>$=$< u_{B \vec{k} z} \, u_{O3 \vec{k} x}>$=$< u_{B \vec{k} z} \, u_{O3 \vec{k} y}>$. A result of the superposition provides the shift of B along the [111] given in the other figure of Fig. 2. A set of the oxygen shifts shows no contradiction with the outcome in the Pb-O case. The rest, positive $< u_{O1 \vec{k} y} \, u_{O2 \vec{k} z}>$, is displayed in Fig. 2 (c), and the equivalencies are given as $< u_{O1 \vec{k} y} \, u_{O2 \vec{k} z}>$=$< u_{O1 \vec{k} z} \, u_{O2 \vec{k} x}>$=$< u_{O1 \vec{k} y} \, u_{O3 \vec{k} x}>$=$< u_{O1 \vec{k} z} \, u_{O3 \vec{k} y}>$=$< u_{O2 \vec{k} x} \, u_{O3 \vec{k} y}>$=$< u_{O2 \vec{k} z} \, u_{O3 \vec{k} x}>$. The resultant schematic shifts are exactly the same as those of the two cases above. In addition, the parameter value for the concentration is also significant only around $\vec{k}$=1/2$<$111$>$ with which oscillation of the concentration, i.e., the 1:1 SRO, is exhibited. But we are not concerned with the SRO in this paper since the concentration correlation has been pretty much studied so far [9,17].

A combination of all the shifts derived from the three main correlations above leads to the local structure shown in Fig. 2 (d). Two pyramids confronting each other whose vertices are occupied by Pb are both translated along the [$\bar{1}\bar{1}\bar{1}$] from the ideal cubic positions given at corners of the shaded box, and maintain point group $\bar{3}$ which the cubic structure also has. On the other hand, the other two B-O pyramids between the two of Pb linking at B realizes the lower symmetry, 3, because of the displacements of B and O. Therefore, this local structure totally forms rhombohedral polar symmetry which was suggested by previous x-ray and neutron 'averaged' crystal structure analysis at low temperature [13]. The similar analysis was also performed by neutron diffuse scattering at single $\vec{k}$ [18] near the zone centers, and the results show general agreement with our local structure. To our knowledge, existence of the polar rhombohedral correlation has crystallographically been proved directly from the diffuse scattering analysis for the first time, and there is no contradiction with results of the NMR work [8].

Quantitative representation of the correlations was done by computing $<u_{s Pb x}\,u_{0 O1 y}>$, $<u_{s B x}\,u_{0 O1 y}>$, and $<u_{s O1 y}\,u_{0 O2 z}>$, according to Eq. (3). They are given in Figures 3 which shows the plots with respect to intercell distance, $\mid\vec{R}_{s}\mid$, along (a) the [111], (b) [110], and (c) [100] directions. For all figures, values of $<u_{s O 1 y}\,u_{0 O2 z}>$ are substantially ten times larger than those of the other two functions. This difference is suggested to originate from phonons which reveal larger displacement amplitudes for lighter elements. Since our x-ray scattering does not distinguish phonons from static displacements, the functions also involve the vibrational displacements. Therefore, the light-element function $<u_{s O1 y}\,u_{0 O2 z}>$ may exhibit the large values compared with the other two. This indistinguishability gives rise to experimental ambiguity on this study somehow.

Fig. 3 (a)-(c) represent continuous modulation of the function values. Portions of the curves with negative $<u_{s Pb x}\,u_{0 O1 y}>$, positive $<u_{s B x}\,u_{0 O1 y}>$, and positive $<u_{s O1 y}\,u_{0 O2 z}>$ (or these with the reverse signs) correspond to the polar correlations as mentioned above. Therefore, lengths of the correlations can read from the figures. In (a)-(c), there are three, four, and three kinds of appreciable peaks of the functions, respectively. Their tail-to-tail separations in sequence of the peak appearance from $\mid\vec{R}_{s}\mid$=0 are $\sim$80, 30, and 30$\AA$ for $\mid\vec{R}_{s}\mid$//[111], $\sim$80, 30, 50, 50$\AA$ for $\mid\vec{R}_{s}\mid$//[110], and $\sim$80, 60, 60$\AA$ for $\mid\vec{R}_{s}\mid$//[100] ($<u_{s O1 y}\,u_{0 O2 z}>$ of the second and third peaks are merged with each other. But because of a dip at the center, we realize that there are two different peaks. The other two peaks are separated clearly.). These data suggest complicated three-dimensional spatial polar modulations. However, we may recognize from the sequence of the lengths that the correlations make spheres or ellipses whose diameters are 30-80$\AA$. These values agree with the length order found in the previous TEM [9] and neutron work [10] which reported that the size is $\sim$100$\AA$ and 30-40$\AA$, respectively. Since TEM measures the length by image contrast, observed diffuse interfaces of the microregions may give this difference. On the other hand, the difference from the neutron result can be attributed to our technical indistinguishability of the dynamic displacements as mentioned above.

Fig. 1 (c) shows the contour map given by the data taken at 270K, corresponding to the map at 300K. The results indicate increase of the intensities near the zone centers (The contours are significantly close to each other making dark areas.) with no essential difference in the distribution from that at 300K, which is also consistent with the previous x-ray work [3]. Similarly, the observed data were analyzed quantitatively, and the recalculated map is represented in Fig. 1 (d). The observed data are well reproduced giving the R factor $\sim$3$\%$. The three displacement-amplitude products, $< u_{Pb \vec{k} x} \, u_{O1 \vec{k} y}>$, $< u_{B \vec{k} x} \, u_{O1 \vec{k} y}>$, and $< u_{O1 \vec{k} y} \, u_{O2 \vec{k} z}>$, are also significantly larger than the others, and construct the same local symmetry as that at 300K. The correlation functions calculated are plotted in Fig. 3 (d)-(f), corresponding to (a)-(c). The results give significant reduction of main-peak values of the functions for the three ionic pairs by a factor of $\sim$1/300, $\sim$1/70, and $\sim$1/1000. This drastic change suggests freezing of thermal lattice fluctuations [3] which may occur near temperature where a dielectric constant peaks, i.e., $\sim$270K. The decrease of the thermal agitation allows new peak structures to appear in the function profiles giving approximately spherical correlations with a diameter of 30-40$\AA$ which is most likely to be a size of the polar microregions. This value is consistent with that reported by the previous neutron work [10].

The data taken at 240K exhibit no significant difference in the intensity distribution from that at 270K. The analysis also gives no change in the local symmetry and its correlation size but shows slight reduction of the function values. This variation may simply be caused by q-dependent freezing [3] or decrease of energy of uncorrelated phonons, and is suggested to be two-phase behavior found by the NMR work [8]. In fact, the function peak structures are more obvious compared with those at 270K.

Below 240K, enlargement of the microregions occurs [9]. Therefore, the diffuse intensities are expected to condense closer to the zone center. This increase must be remarkable in our unaccessible zone-center region, $\sim$$(0.01)^{3}$$r.l.u.^{3}$, and was obstruction to perform the measurement at the lower temperature. The higher instrumental resolution is required for the further investigation.

The polar domains which appear through the local condensation exhibit the local polarizations randomly along anyone of the $<$111$>$ directions giving the macroscopic paraelectric state, and the polarizations must be in a sort of frustrating states under an electric field, which leads to the dielectric dispersion. A zero-frequency picture of the random polarizations gives an idea that the Coulombic field interactions among the microdomains have a tendency to prevent their growth each other from the embryonic states. This effect requires undercooling to promote the further transition. The TEM study [9] and another previous synchrotron x-ray work [19] determined that temperature of the initiation is 220-230K, meaning undercooling of several-ten kelvins. Local lattice distortion generated by the charge misfit and the size effect simply provides the preferential condensation sites heterogeneously, and causes a broad maximum of the dielectric constant. The dispersion is just a consequence of the electric-energetical growth barrier.

N.T. thanks Y. Uesu, K. Fujishiro, N. Yamamoto, and G. Shirane for valuable private communications on this study. The acknowledgment is extended to K. Ohshima, K. Ohwada, and Y. Katsuki for their assistance in the measurement.

\begin{figure}
\caption{The 0th-(100) contour maps of (a) observed $I(\vec{q}_{1})$ at 300K and (b) its calculated result, and correspondingly (c) and (d) at 270K. The contour values are indicated in $10^{4}$ e.u. for all pieces. The contour interval is 0.8 in the units.}
\end{figure}
\begin{figure}
\caption{Schematic representation of (a) $< u_{Pb \vec{k} x} \, u_{O1 \vec{k} y}>$, (b) $< u_{B \vec{k} x} \, u_{O1 \vec{k} y}>$, (c) $< u_{O1 \vec{k} x} \, u_{O2 \vec{k} y}>$, and their resultant shifts given in terms of the symmetry Pm3m. Superposition of all ionic shifts shows the correlation which is rhombohedral polar as illustrated in (d), and the symmetry corresponds to R3m.}
\end{figure}
\begin{figure}
\caption{The correlation functions of intercell distance $\vec{R}_{s}$ at 300K along (a) the [111], (b) [110], and (c) [100] directions. Figures of the functions at 270K corresponding to (a)-(c) are given as (d)-(f). Values of the three functions, $<u_{s Pb x}\,u_{0 O1 y}>$, $<u_{s B x}\,u_{0 O1 y}>$, and $<u_{s O1 y}\,u_{0 O2 z}>$, are indicated in each figure by circles, squares, and triangles, respectively.}
\end{figure}


\begin{references}
\bibitem{1} G. A. Smolenskii, V. A. Isupov, A. I. Agranovskaya, and S. N. Popov, Sov. Phys. Solid State {\bf 2}, 2584 (1961).
\bibitem{2} L. E. Cross, Ferroelectrics {\bf 76}, 241 (1987).
\bibitem{3} H. You and Q. M. Zhang, Phys. Rev. Lett. {\bf 79}, 3950 (1997).
\bibitem{4} N. Takesue, Y. Fujii, M. Ichihara, and H. Chen, Phys. Rev. Lett. {\bf 82}, 3709 (1999).
\bibitem{5} N. Takesue, Y. Fujii, M. Ichihara, H. Chen, S. Tatemori, and J. Hatano, J. of Phys. Cond. Matt. {\bf 11}, 8301 (1999).
\bibitem{6} V. Westphal, W. Kleemann, and M. D. Glinchuk, Phys. Rev. Lett. {\bf 68}, 847 (1992).
\bibitem{7} E. V. Colla, E. Yu. Koroleva, N. M. Okuneva, and S. B. Vakhrushev, Phys. Rev. Lett. {\bf 74}, 1681 (1995).
\bibitem{8} R. Blinc, J. Dolins$\check{e}$k, A. Gregorovi$\check{c}$, B. Zalar, C. Filipi$\check{c}$, Z. Kutnjak, A. Levstik, and R. Pirc, Phys. Rev. Lett. {\bf 83}, 424 (1999).
\bibitem{9} M. Yoshida, S. Mori, N. Yamamoto, Y. Uesu, and J. M. Kiat, Ferroelectrics {\bf 217}, 327 (1998).
\bibitem{10} P. M. Gehring, S.-E. Park, G. Shirane, Phys. Rev. Lett. {\bf 84}, 5216 (2000).
\bibitem{11} M. A. Krivoglatz, {\it Theory of X-Ray and Thermal-Neutron Scattering by Real Crystals}, (Plenum Press, New York, 1969), Chapter 8.
\bibitem{12} {\it International Tables for X-Ray Crystallography} {\bf 4}, published for The International Union of Crystallography, (The Kynoch Press, Birmingham, England, 1974).
\bibitem{13} P. Bonneau, P. Garnier, G. Calvarin, E. Husson, J. R. Gavarri, A. W. Hewat, and A. Morell, J. Sol. Stat. Chem. {\bf 91}, 350 (1991).
\bibitem{14} L. H. Schwartz and J. B. Cohen, {\it Diffraction from Materials}, 2nd ed., (Springer-Verlag, Berlin, 1987), Appendix C.
\bibitem{15} D. T. Cromer and J. B. Mann, J. of Chem. Phys. {\bf 47}, 1892 (1967); D. T. Cromer, J. of Chem. Phys. {\bf 50}, 4857 (1969).
\bibitem{16} N. Takesue and K. Hirota, private communication.
\bibitem{17} Q. M. Zhang and H. You, Sol. Stat. Commun. {\bf 97}, 693 (1996).
\bibitem{18} S. B. Vakhrushev, A. A. Naberezhnov, N. M. Okuneva, and B. N. Savenko, Phys. Solid State {\bf 37}, 1993 (1995).
\bibitem{19} S. Vakhrushev, A. Naberezhnov, S. K. Shinha, Y. P. Peng, and T. Egami, J. Phys. Chem. Sol. {\bf 57}, 1517 (1996).
\end{references}
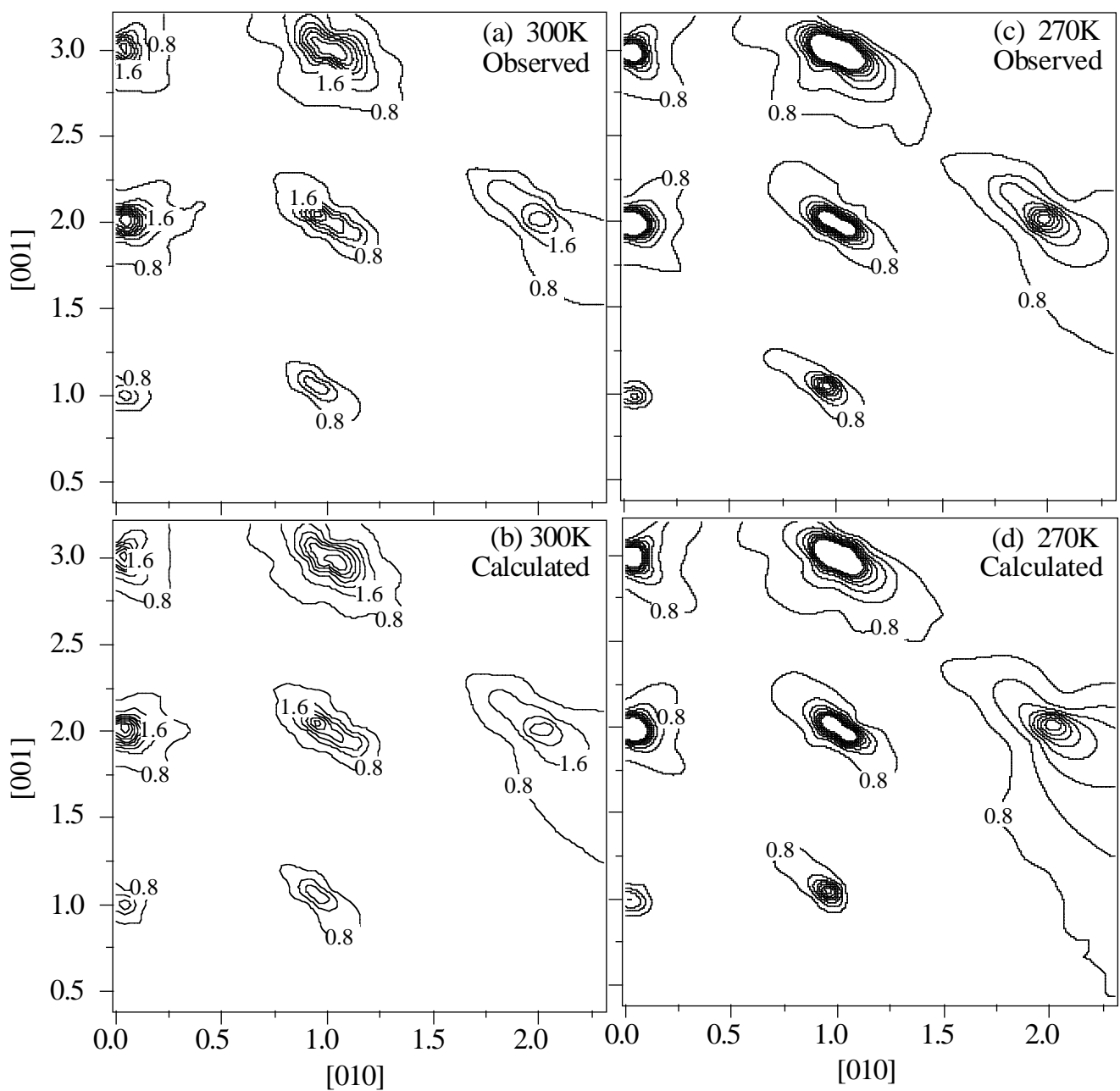
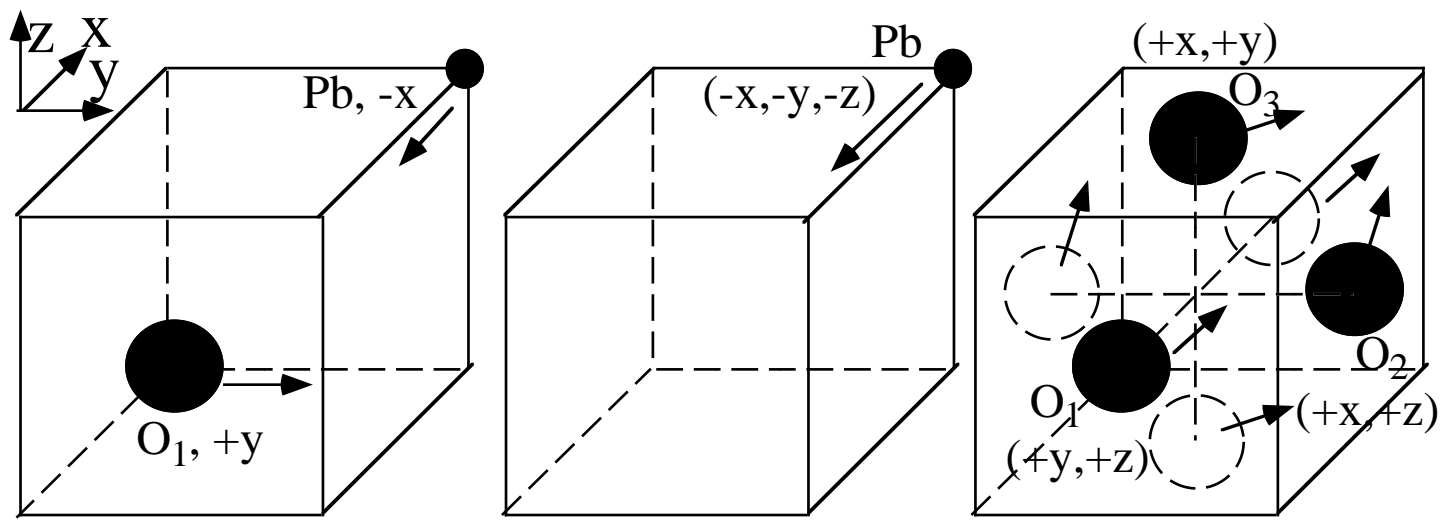
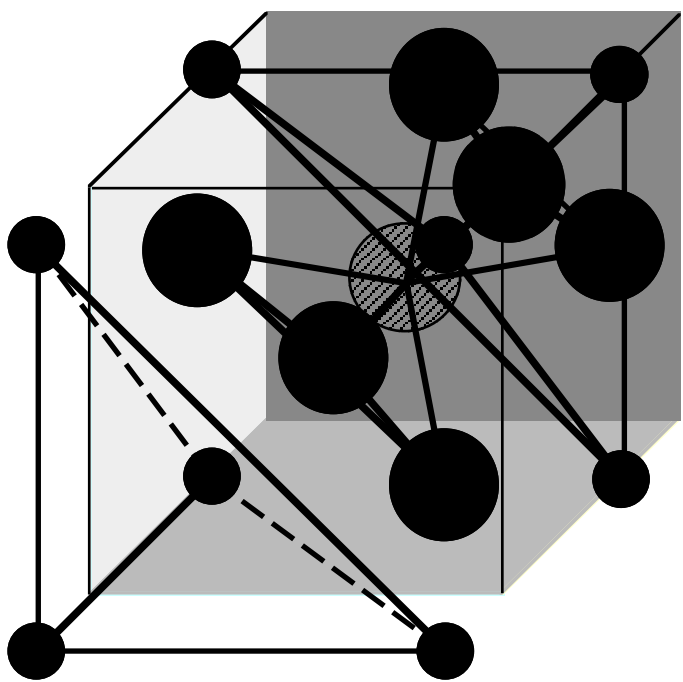
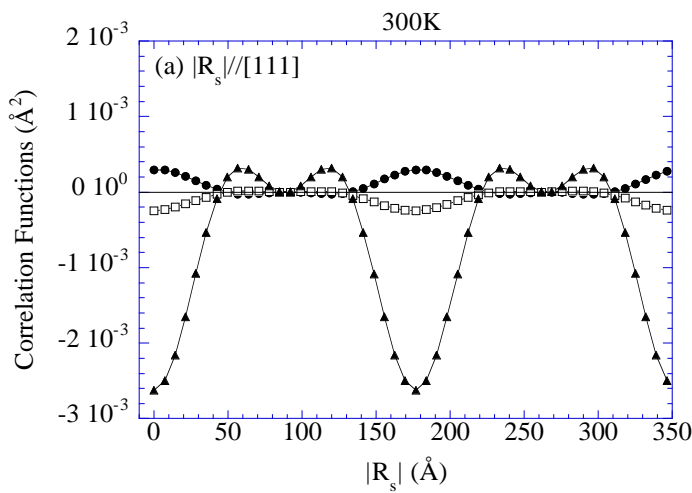
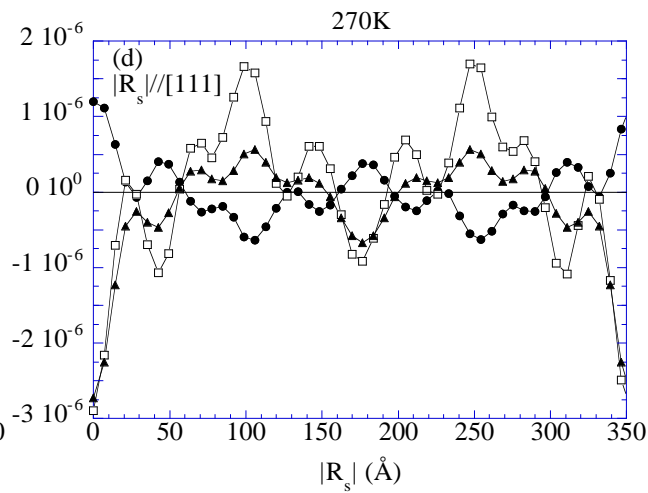
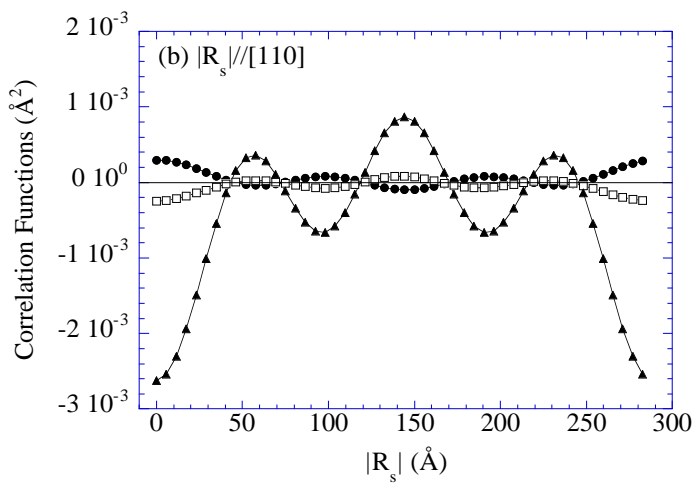
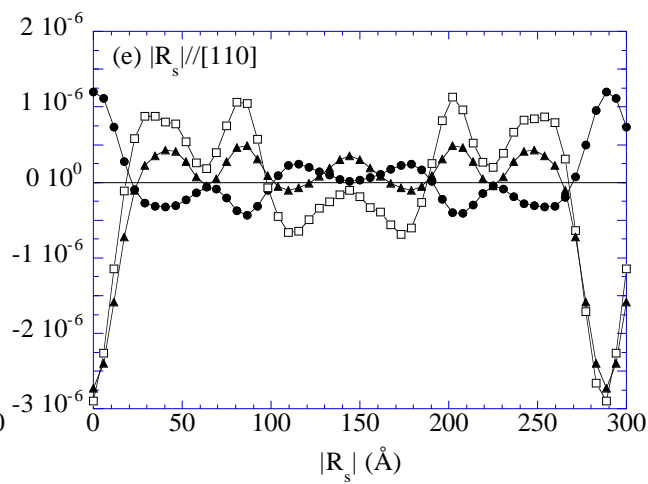
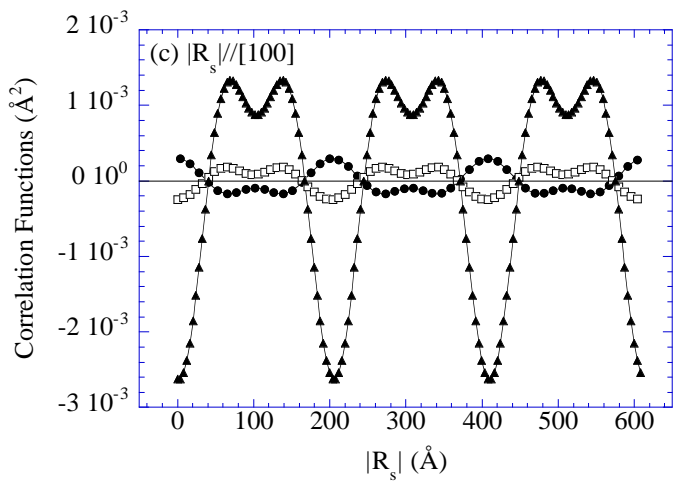
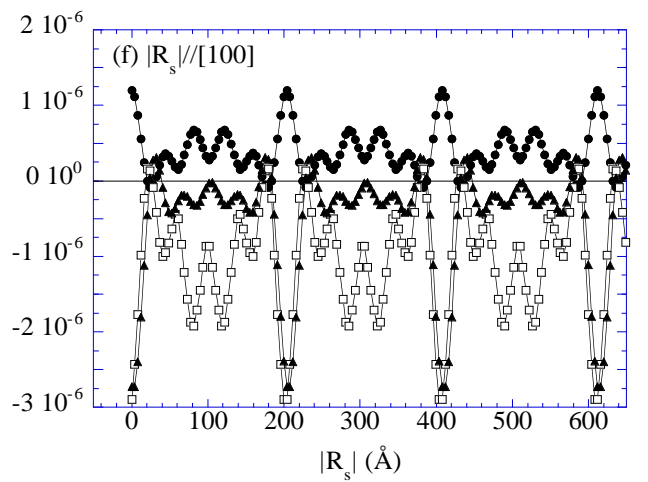
\end{document}